\documentclass{article}
\usepackage{spconf,amsmath,amsthm,amssymb,graphicx}

\usepackage{pgf,tikz,pgfplots}
\usepackage{mathrsfs}
\usetikzlibrary{arrows}
\usepackage{mathtools} 
\usepackage{hyperref}
\hypersetup{
	colorlinks=true,
	linkcolor=blue,
	filecolor=magenta,
	urlcolor=cyan,
	citecolor=blue
}
\usepackage{subfigure}
\usepackage{ifthen}
\usepackage[utf8]{inputenc} 
\usepackage[T1]{fontenc}
\usepackage{caption}
\usepackage{algorithm,algorithmic}
\newtheorem{theorem}{\bf Theorem}

\usepackage{mathtools}
\usepackage{cuted}

\usepackage{lipsum,graphicx,multicol}

\usepackage{cases}

\usepackage{float}

\title{Optimal QoS-Aware Network Slicing for Service-Oriented Networks with Flexible Routing}
\name{Wei-Kun Chen$^{\star}$, Ya-Feng Liu$^{\dag}$	
Yu-Hong Dai$^{\dag}$, and Zhi-Quan Luo$^\ddag$
}
\address{ 
	$^{\star}$School of Mathematics and Statistics, Beijing Institute of Technology, Beijing, China\\[2pt]
	$^{\dag}$LSEC, ICMSEC, AMSS, Chinese Academy of Sciences, Beijing, China\\[2pt]
	$^{\ddag}$Shenzhen Research Institute of Big Data and The Chinese University of Hong Kong, Shenzhen, China\\[2pt]
	Email:
	chenweikun@bit.edu.cn, \{yafliu,~dyh\}@lsec.cc.ac.cn, luozq@cuhk.edu.cn
}

\begin{document}
%
\ninept
\maketitle
\setlength{\abovedisplayskip}{0.09cm}
\setlength{\belowdisplayskip}{0.09cm}
\setlength{\jot}{0.09cm}
\begin{abstract}
	In this paper, we consider the network slicing problem which attempts to map multiple customized virtual network requests (also called services) to a common shared network infrastructure and allocate network resources to meet diverse quality of service (QoS) requirements.
	We first propose a mixed integer nonlinear program (MINLP)  formulation for this problem that optimizes the network resource consumption while jointly considers QoS requirements, flow routing, and resource budget constraints.
	In particular, the proposed formulation is able to flexibly route the traffic flow of the services on multiple paths and provide end-to-end (E2E) delay and reliability guarantees for all services.
	Due to the intrinsic nonlinearity, the MINLP formulation is computationally difficult to solve. 
	To overcome this difficulty, we then propose a mixed integer \emph{linear} program (MILP) formulation and show that the two formulations and their continuous relaxations are equivalent.
	Different from the continuous relaxation of the MINLP formulation which is a nonconvex nonlinear programming problem, the continuous relaxation of the MILP formulation is a polynomial time solvable linear programming problem, which makes the MILP formulation much more computationally solvable.
	Numerical results demonstrate the effectiveness and efficiency of the proposed formulations over existing ones.
\end{abstract}
\section{Introduction}
\label{sec:intro}

Network function virtualization (NFV) is one of the key technologies for the fifth generation (5G) and beyond 5G (B5G) networks \cite{Mijumbi2016}.
Different from traditional networks in which service requests (e.g., high dimensional video, virtual private network, and remote robotic surgery) are implemented by dedicated hardware in fixed locations, NFV-enabled networks efficiently leverage virtualization technologies to configure some specific cloud nodes in the network to process network service functions on-demand, and then establish a customized virtual network for all service requests.
However, since virtual network functions (VNFs) of all services are implemented over a single shared cloud network infrastructure, it is vital to efficiently allocate network (e.g., cloud and communication) resources subject to diverse quality of service (QoS) requirements (e.g., E2E delay and reliability requirements) of all services and the capacity constraints of all cloud nodes and links in the network.

We call the above resource allocation problem in the NFV-enabled network \emph{network slicing}, which jointly considers the VNF placement (that maps VNFs into cloud nodes in the network) and the traffic routing (that finds paths connecting two cloud nodes which perform two adjacent VNFs in the network).
In recent years, there are considerable works on network slicing and its variants; see \cite{Zhang2017}-\cite{Yeow2010} and the references therein.
In particular, references \cite{Zhang2017}-\cite{Mijumbi2015} considered the network slicing problem with a limited network resource constraint but neither considered E2E delay nor E2E reliability constraints of the services.
References \cite{Chen2021a}-\cite{Addis2015}  
incorporated the E2E delay constraint of the services into their formulations but still did not take the E2E reliability constraint of the services into consideration.
Obviously, formulations that do not consider E2E delay or E2E reliability constraints may return a solution that does not satisfy these QoS requirements.
References \cite{Jiang2012}-\cite{Guo2011} and 
\cite{Vizarreta2017}-\cite{Mohammadkhan2015}
simplified the traffic routing strategy
by selecting paths from a predetermined path set and enforcing a single path routing, respectively.
Apparently, formulations based on such assumptions do not fully exploit the flexibility of traffic routing, thereby leading to a poor performance of the whole network.
References \cite{Qu2017}-\cite{Yeow2010}
proposed different protection schemes (which reserve additional cloud node or link capacities to provide resiliency against links' or  cloud nodes' failure) to guarantee the E2E reliability of the services.
However, the protection schemes generally lead to inefficiency of resource allocation, as it requires more link or node capacity consumption.

To summarize, for the network slicing problem, none of the existing formulations/works  simultaneously takes all of the above practical factors (e.g., flexible routing, E2E delay and reliability, and network resource consumption) into consideration.
The goal of this work is to provide a mathematical formulation of the network slicing problem that simultaneously allows the traffic flows to be flexibly transmitted on (possibly) multiple paths, satisfies the E2E delay and reliability requirements of all services, and considers the network resource consumption.
In particular, we first formulate the problem as a mixed integer {nonlinear} program (MINLP) which minimizes the network resource consumption subject to the capacity constraints of all cloud nodes and links and the E2E delay and reliability constraints of all services.
Due to the intrinsic nonlinearity, the MINLP formulation appears to be computationally difficult to solve. 
To overcome this difficulty, we then propose an equivalent mixed integer \emph{linear} program (MILP) formulation and show that the continuous relaxations of the two formulations are equivalent (in terms of sharing the same optimal solutions).
However, in sharp contrast to the continuous relaxation of the MINLP formulation which is a nonconvex nonlinear programming (NLP) problem, the continuous relaxation of the MILP formulation is a polynomial time solvable linear programming (LP) problem, which makes the MILP formulation much more computationally solvable.
The key of transforming MINLP into MILP is a novel way of rewriting \emph{nonlinear} flow conservation constraints for traffic routing as \emph{linear} constraints, which is a technical contribution of this paper.
Finally, numerical results demonstrate the effectiveness and efficiency of our proposed formulations over existing ones.

\section{System model and problem formulation}
\label{sec:modelformulation}
We use $\mathcal{G}=\{\mathcal{I},\mathcal{L}\}$ to represent the substrate (directed) network, where $\mathcal{I}=\{i\}$ and $\mathcal{L}=\{(i,j)\}$ denote the sets of nodes and links, respectively. 
Let $ \mathcal{V} \subseteq \mathcal{I}$ be the set of cloud nodes.
Each cloud node $ v $ has a computational capacity $ \mu_v $ and a reliability $\gamma_{v}$ \cite{Vizarreta2017,Guerzoni2014}.
As assumed in \cite{Zhang2017}, processing one unit of data rate consumes one unit of (normalized) computational capacity.
Each link $ (i,j) $ has an expected (communication) delay $ d_{ij} $ \cite{Woldeyohannes2018}, a reliability $\gamma_{ij}$ \cite{Vizarreta2017,Guerzoni2014}, and  a total data rate upper bounded by the capacity $C_{ij}$.
A set of services $\mathcal{K}=\{k\}$ is needed to be supported by the network.
Let $S(k),D(k)\notin \mathcal{V}$ be the source and destination nodes of service $k$.
Each service $ k $ relates to a customized service function chain (SFC) consisting of $ \ell_k $ service functions that have to be processed in sequence by the network: $f_{1}^k\rightarrow f_{2}^k\rightarrow \cdots \rightarrow f_{\ell_k}^k$ \cite{Zhang2013,Halpern2015,Mirjalily2018}.
As required in \cite{Zhang2017,Woldeyohannes2018}, in order to minimize the coordination overhead, each function must be processed at exactly one cloud node.
If function $ f^k_s $, $ s \in \mathcal{F}(k) := \{1,\ldots, \ell_k\} $, is processed by cloud node $ v $ in $ \mathcal{V} $, the expected NFV delay is assumed to be known as $d_{v,s}(k) $, which includes both processing and queuing delays \cite{Luizelli2015,Woldeyohannes2018}.
For service $ k $, let $ \lambda^k_0 $ and $ \lambda^k_s $ denote the data rates before receiving any function and after receiving {function} $ f^k_s $, respectively.
Each service $ k $ has an E2E delay requirement and an E2E reliability requirement, denoted as $ \Theta_k $ and $\Gamma_k$, respectively.

The network slicing problem is to determine VNF placement, the routes, and the associated data rates on the corresponding routes of all services while satisfying the capacity constraints of all cloud nodes and links and the E2E delay and reliability constraints of all services.
Next, we shall present the constraints and objective function of the problem formulation in details.
\\[2pt]
{\bf\noindent$\bullet$ VNF Placement\\[2pt]}
\indent Let $x_{v,s}(k)=1$ indicate that function $f^k_s$ is processed by cloud node $v$; otherwise, $x_{v,s}(k)=0$.
Each function $f_s^k$ must be processed by exactly one cloud node, i.e.,
\begin{eqnarray}
\label{onlyonenode}
\sum_{v\in \mathcal{V}}x_{v,s}(k)=1,~\forall ~k \in \mathcal{K},~ \forall ~s\in  \mathcal{F}(k).
\end{eqnarray}
In addition, we introduce binary variable $x_v(k)$ to denote whether there exists at least one function in $\mathcal{F}(k)$ processed by cloud node $v$ and binary variable $y_v$ to denote whether cloud node $v$ is activated and powered on.
By definition, we have 
\begin{align}
\label{xyxelation}
& x_{v,s}(k) \leq  x_{v}(k), ~ \forall~v \in \mathcal{V},~\forall~k \in \mathcal{K},~\forall~s \in \mathcal{F}(k), \\
& x_{v}(k) \leq  y_v, ~ \forall~v \in \mathcal{V},~\forall~k \in \mathcal{K}.\label{xyrelation}
\end{align}
The node capacity constraints can be written as follows:
\begin{equation}
\label{nodecapcons}
\sum_{k\in \mathcal{K}}\sum_{s \in \mathcal{F}(k)}\lambda_s^k x_{v,s}(k)\leq \mu_v y_v,~\forall~ v \in \mathcal{V}.
\end{equation}
{\bf\noindent$\bullet$ Traffic Routing\vspace{0.1cm}\\}
\indent Let $ (k,s) $ denote the traffic flow which is routed between the two cloud nodes hosting the two adjacent functions $ f_s^k $ and $ f_{s+1}^k $.
We follow \cite{Chen2021a} to assume that there are at most $P$ paths that can be used to route flow $(k,s)$ and denote $\mathcal{P}=\{1, \ldots, P\}$.
%
%
Let binary variable $ z_{ij}(k,s,p)$ denote whether (or not) link $ (i,j) $ is on the $ p $-th path of flow $ (k,s) $.
Then, to ensure that the functions of each service $k$ are processed in the prespecified order $f_{1}^k\rightarrow f_{2}^k\rightarrow \cdots \rightarrow f_{\ell_k}^k$, we need
\begin{align}
& \sum_{j: (j,i) \in \mathcal{{L}}} z_{ji}(k, s, p) - \sum_{j: (i,j) \in \mathcal{{L}}} z_{ij}(k, s,  p)=b_{i,s}(k),\nonumber\\[-2pt]
& ~~~\qquad \qquad~~\forall~ i \in \mathcal{I},~\forall~k \in \mathcal{K},~\forall~s \in \mathcal{F}(k)\cup \{0\},~ \forall~ p \in \mathcal{P}, \label{SFC1}
\end{align}
where
\begin{equation*}
b_{i,s}(k)=\left\{\begin{array}{ll}
\!\!\!\!-1,&\!\!\!\!\text{if~}s=0~\text{and}~i= S(k);\\
\!\!\!\!x_{i,s+1}(k),&\!\!\!\!\text{if~}s=0~\text{and}~i\in \mathcal{V};\\
\!\!\!\!x_{i,s+1}(k)-x_{i,s}(k),&\!\!\!\!\text{if~} s\in \{1, \ldots,\ell_k-1\}
~\text{and}~i\in \mathcal{V};\\
\!\!\!\!-x_{i,s}(k),&\!\!\!\!\text{if~}s=\ell_k
~\text{and}~i\in \mathcal{V};\\
\!\!\!\!1,&\!\!\!\!\text{if~}s=\ell_k
~\text{and}~i=D(k);\\
\!\!\!\!0,& \!\!\!\!\text{otherwise}.
\end{array}
\right.
\end{equation*}


Next, we follow \cite{Chen2021b,Promwongsa2020} to present the flow conservation constraints for the data rates.
To proceed, we need variable  $ r(k,s,p)\in [0,1] $ denoting the fraction of data rate $\lambda_s^k$ on the $ p $-th path of flow $ (k,s) $ and variable   $ r_{ij}(k,s,p) \in [0,1]$ denoting the fraction of data rate $\lambda_{s}^k$ on link $(i,j)$ (when $z_{ij}(k,s,p)=1$).
By definition, we have
\begin{align}
	& \sum_{p \in \mathcal{P}}  r(k, {s}, p) =  1,   ~ \forall~ k \in \mathcal{K},~\forall~s\in \mathcal{F}(k)\cup \{0\} \label{relalambdaandx11},\\[-2pt]
	& r_{ij}(k, s,  p ) = r(k,s,p) z_{ij}(k, s,p ), \nonumber                                                                                                  \\
	& \qquad\quad~ \forall~(i,j) \in {\mathcal{L}}, ~\forall~k \in \mathcal{K}, ~\forall~s \in \mathcal{F}(k)\cup \{0\},~\forall~p \in \mathcal{P}. \label{nonlinearcons}
\end{align}
Note that constraint \eqref{nonlinearcons} is nonlinear.
Finally, the total data rates on link $ (i,j) $ is upper bounded by capacity $ C_{ij} $:
\begin{equation}
\label{linkcapcons1}
\sum_{k \in \mathcal{K}} \sum_{s\in \mathcal{F}(k) \cup \{0\}}\sum_{p \in \mathcal{P}} \lambda_{s}^k r_{ij}(k, s,p) \leq C_{ij}, ~  \forall~(i,j) \in \mathcal{L} .
\end{equation}
{\bf\noindent$\bullet$ E2E Reliability\\[2pt]}
\indent To model the reliability of each service, we introduce binary variable $z_{ij}(k)$ denoting whether link $(i,j)$ is used to route the traffic flow of service $k$.
By definition, we have 
\begin{align}
& z_{ij}(k, s,  p ) \leq z_{ij}(k), \nonumber                                                                                          \\
& \qquad\quad~ \forall~(i,j) \in {\mathcal{L}}, ~\forall~k \in \mathcal{K}, ~\forall~s \in \mathcal{F}(k)\cup \{0\},~\forall~p \in \mathcal{P}. \label{zzrelcons}
\end{align}
The E2E reliability of service $k$ is defined as the product of the reliabilities of all cloud nodes hosting all functions $f_s^k$ in its SFC, and the reliabilities of all links used by service $k$ \cite{Vizarreta2017}.
The following constraint ensures that the E2E reliability of service $k$ is larger than or equal to its given threshold $\Gamma_k\in[0,1]$:
\begin{equation}
	\label{E2Ereliability1}
	\tag{$\star$}
	\prod_{v\in \mathcal{V}} \rho_{v}(k)\cdot 	\prod_{(i,j)\in \mathcal{L}} \rho_{ij}(k) \geq \Gamma_k,~\forall~k \in \mathcal{K},
\end{equation}
where 
\begin{equation*}
\begin{aligned}
\rho_{v}(k)=\left\{\begin{array}{ll}
\gamma_v,&\text{if~}x_{v}(k)=1;\\
1,&\text{if~}x_{v}(k)=0;
\end{array}
\right. & 
\rho_{ij}(k)=\left\{\begin{array}{ll}
\gamma_{ij},&\text{if~}z_{ij}(k)=1;\\
1,&\text{if~}z_{ij}(k)=0.
\end{array}
\right.
\end{aligned}
\end{equation*}
The above nonlinear constraint \eqref{E2Ereliability1} can be equivalently linearized as follows.
By the definitions of $\rho_v(k)$ and $\rho_{ij}(k)$, we have
\begin{equation*}
\log (\rho_{v}(k))= \log (\gamma_v) \cdot x_{v}(k)  ~\text{and}~ \log (\rho_{ij}(k))=\log (\gamma_{ij}) \cdot z_{ij}(k).
\end{equation*}
Then we can apply the logarithmic transformation on both sides of \eqref{E2Ereliability1} and obtain an equivalent \emph{linear} E2E reliability constraint: 
\begin{equation}\label{E2Ereliability2}
\sum_{v\in \mathcal{V}}  \log (\gamma_v) \cdot x_{v}(k) +  	\sum_{(i,j)\in \mathcal{L}} \log (\gamma_{ij}) \cdot z_{ij}(k)\geq \log(\Gamma_k),~\forall~k \in \mathcal{K}.
\end{equation}
%
{\bf\noindent$\bullet$ E2E delay\vspace{0.1cm}\\}
\indent We use variable $ \theta(k,s) $ to denote the communication delay due to the traffic flow from the cloud node hosting function $ f^k_s $ to the cloud node hosting function $ f^k_{s+1} $. 
By definition, we have
\begingroup
\allowdisplaybreaks
\begin{align}
& \theta(k,s) \geq \sum_{(i,j) \in \mathcal{{L}}}  d_{ij}  z_{ij}(k, s, p),\nonumber                                                   \\[-2pt]
& \qquad\qquad\qquad\qquad~~  \forall~k \in \mathcal{K}, ~ \forall~s \in \mathcal{F}(k) \cup \{0\},~\forall ~p \in \mathcal{P} \label{consdelay2funs1}.
\end{align}
\endgroup 
To guarantee that service $k$'s {E2E} delay is not larger than its threshold $\Theta_k$, we need the following constraint:
\begin{equation}
\label{delayconstraint}
\theta_N(k) +\theta_L(k)  \leq \Theta_k,~\forall~k \in  \mathcal{K},
\end{equation}
where  $\theta_N(k) =  \sum_{v \in \mathcal{{V}}}\sum_{s \in \mathcal{F}(k)} d_{v,s}(k) x_{v,s}(k)$ and $\theta_L(k) = \sum_{s \in \mathcal{F}(k)\cup \{0\}} \theta(k,s)$ denote the total NFV delay on the nodes and the total communication delay on the links of service $ k $, respectively.\\[2pt]
{\bf\noindent$\bullet$ Problem Formulation\\[2pt]}
\indent The network slicing problem is to minimize a weighted sum of the total number of activated nodes (equivalent to the total power consumption in the cloud network \cite{Chen2021a}) and the total link capacity consumption in the whole network:
\begin{align}
& \min_{\substack{\boldsymbol{x},\boldsymbol{y},\\\boldsymbol{r},\boldsymbol{z},\boldsymbol{\theta}}} &  & \sum_{v \in \mathcal{V}}y_v + \sigma \sum_{(i,j)\in \mathcal{L}} \sum_{k\in \mathcal{K}} \sum_{s\in \mathcal{F}(k)\cup \{0\}} \sum_{p\in \mathcal{P}} r_{ij}(k,s,p) \nonumber \\
& ~~{\text{s.\,t.~}}                                                                    &  & \eqref{onlyonenode}-\eqref{delayconstraint}, \nonumber \\
& &&  x_{v,s}(k),~x_v(k),~y_v\in\{0,1\},\,\forall \,k\in\mathcal{K}, ~s\in \mathcal{F}(k),~v\in\mathcal{{V}}, \nonumber\\
& && r(k,s,p),~r_{ij}(k, s, p )\geq 0,~z_{ij}(k, s, p ),~z_{ij}(k)\in \{0,1\}, \nonumber\\
& && \qquad \qquad~~~\forall~(i,j)\in \mathcal{L},~k\in \mathcal{K},~s\in \mathcal{F}(k)\cup \{0\},~p \in \mathcal{P}, \nonumber\\
& &&  \theta(k,s)\geq 0,~\forall~k \in \mathcal{K},~s \in \mathcal{F}(k)\cup \{0\},
\label{minlp}
\tag{\text{MINLP}}
\end{align}
where $ \sigma $ is a constant that balances the two terms in the objective function.
The technical advantage of incorporating the (second) link capacity consumption term into the objective function of formulation \eqref{minlp} is as follows.
First, it is helpful in avoiding cycles in the traffic flow between the two nodes hosting the two adjacent service functions.
Second, minimizing the total link capacity consumption enables an efficient reservation of more link capacities for future use \cite{Woldeyohannes2018}.
Third, minimizing the total data rate in the whole network further decreases the total E2E delay and reliability of all services.

It is worth remarking that formulation \eqref{minlp} can be reformulated as an MILP formulation, as nonlinear constraint \eqref{nonlinearcons} can be equivalently linearized as
\begin{equation}
\label{bilinearcons}
\tag{7'}
\begin{aligned}
& r_{ij}(k,s,p) \geq  z_{ij}(k, s,p) + r(k,s,p) -1,   
\\
& r_{ij}(k,s,p) \leq   z_{ij}(k, s,p) ,~ r_{ij}(k,s,p) \leq  r(k,s,p).                                   
\end{aligned}
\end{equation}
However, the above linearization generally leads to a weak continuous relaxation (i.e., relaxing all binary variables to continuous variables in $[0,1]$) \cite{Chen2021c}.
As such, it is inefficient to  employ a standard solver in solving \eqref{minlp}, as demonstrated in our experiments.
In the next section, we shall present another equivalent MILP formulation that is much more computationally solvable.

Finally, we would like to highlight the difference between our proposed formulation \eqref{minlp} and the two closely related works \cite{Vizarreta2017} and \cite{Chen2021b}.
More specifically, different from that in \cite{Vizarreta2017} where a single path routing strategy is used for the traffic flow of each service (between two cloud nodes hosting two adjacent functions of a service), our proposed formulation allows to transmit the traffic flow of each service on (possibly) multiple paths and hence fully exploits the flexibility of traffic routing; in sharp contrast to that in \cite{Chen2021b}, our proposed formulation can guarantee the E2E reliability of all services.
\section{A novel MILP problem formulation}

In this section, we shall derive a novel MILP formulation for the network slicing problem which is mathematically equivalent to formulation \eqref{minlp} but much more computationally solvable.\\[2pt]
%
{\bf\noindent$\bullet$New Linear Flow Conservation Constraints for Traffic Routing\\[2pt]}
\indent Recall that in the previous section, we use \eqref{relalambdaandx11} and nonlinear constraint \eqref{nonlinearcons} to ensure the flow conservation for the data rates of the $p$-th path of flow $(k,s)$.
However, the intrinsic nonlinearity in it leads to an inefficient solution of formulation \eqref{minlp}.
To overcome this difficulty, we reformulate \eqref{relalambdaandx11}-\eqref{nonlinearcons} as  
\begin{align}
& \sum_{j: (i,j) \in \mathcal{{L}}} z_{ij}(k, s,  p)\leq 1, \nonumber\\[-3pt]
& ~\qquad \qquad~~\forall~i \in \mathcal{I},~\forall~k \in \mathcal{K},~\forall~s \in \mathcal{F}(k)\cup \{0\},~ \forall~ p \in \mathcal{P}, \label{SFC0}\\
& r_{ij}(k, s,  p ) \leq z_{ij}(k, s,p ), \nonumber                                                                                          \\
& \qquad~~~\forall~(i,j) \in {\mathcal{L}}, ~\forall~k \in \mathcal{K}, ~\forall~s \in \mathcal{F}(k)\cup \{0\},~\forall~p \in \mathcal{P}, \label{linearcons}\\
& \sum_{p \in \mathcal{P}}\sum_{j: (j,i) \in \mathcal{{L}}} r_{ji}(k, s, p) - \sum_{p \in \mathcal{P}}\sum_{j: (i,j) \in \mathcal{{L}}} r_{ij}(k, s,  p)= b_{i,s}(k),\nonumber\\
& \qquad \qquad\qquad\qquad\qquad\qquad\qquad \forall~k \in \mathcal{K}, ~(s,i) \in \mathcal{SI}(k), \label{SFC2}
\end{align}
and \eqref{SFC3}-\eqref{SFC5} on top of next page.
\begin{figure*}[t]
	\begin{numcases}{\sum_{j: (j,i) \in \mathcal{{L}}} r_{ji}(k, s, p) -\sum_{j: (i,j) \in \mathcal{{L}}} r_{ij}(k, s,  p)}
	=0,&$\forall~k \in \mathcal{K},~\forall~s\in \mathcal{F}(k)\cup \{0\},~\forall~i\in \mathcal{I}\backslash\mathcal{V},~\forall~p \in \mathcal{P}$;\label{SFC3}\\
	\leq x_{i,s+1}(k),&$\forall~k \in \mathcal{K},~\forall~s\in \mathcal{F}(k)\cup \{0\}\backslash\{\ell_k\},~\forall~i\in \mathcal{V},~\forall~p \in \mathcal{P}$;\label{SFC4}\\
	\geq -x_{i,s}(k),&$\forall~k \in \mathcal{K},~\forall~s\in \mathcal{F}(k),~\forall~i\in \mathcal{V},~\forall~p \in \mathcal{P}$\label{SFC5}.
	\end{numcases}
	\hrulefill
	\vspace*{-18pt}
\end{figure*}
Here 
\begin{align*}
&\mathcal{SI}(k) \!=\!  \left \{ (0,i)  :  i \in \mathcal{V}\cup \{S(k)\} \right \}  \cup  \left\{ (\ell_k,i)  :  i \in \mathcal{V}\cup \{D(k)\} \right\} \\
&  \qquad \qquad \qquad \qquad\qquad~ \cup  \left\{ (s,i) :  s\in \{1,\ldots, \ell_k-1\},~i\in \mathcal{V} \right\}.
\end{align*}
\noindent Constraint \eqref{SFC0} ensures that there exists at most one link leaving node $i$ for the $p$-th path of flow $(k,s)$.
Constraint \eqref{linearcons} requires that if $r_{ij}(k,s,p)> 0$, $z_{ij}(k,s,p)=1$ must hold.
Constraint \eqref{SFC2} ensures the flow conservation of data rates at the source and destination of flow $(k,s)$.
More specifically, it guarantees that (i) the total fraction of data rate $\lambda_{s}^k$ leaving the source node $S(k)$ $(s=0)$ and the node hosting function $f_s^k$ ($s\in \{1,\ldots, \ell_k \}$), and (ii) the total fraction of data rate $\lambda_{s}^k$ entering into the node hosting function $f_{s+1}^k$ ($s\in \{0, \ldots,\ell_k -1\}$) and the destination node $D(k)$ ($s=\ell_k$) are all equal to $1$.
Finally, constraints \eqref{SFC3}-\eqref{SFC5} ensure the flow conservation of the data rate at each intermediate node of the $p$-th path of flow $(k,s)$. 
Notice that when $x_{i,s}(k)=x_{i,s+1}(k)=0$ for some $i\in \mathcal{V}$, cloud node $i$ is also an intermediate node of flow $(k,s)$, and in this case, combining \eqref{SFC4} and \eqref{SFC5}, we also have $ \sum_{j: (j,i) \in \mathcal{{L}}} r_{ji}(k, s, p) -\sum_{j: (i,j) \in \mathcal{{L}}} r_{ij}(k, s,  p)=0 $.
Note that in the new formulation, we do not need variables $\{r(k,s,p)\}$ and all constraints \eqref{SFC0}-\eqref{SFC5} are \emph{linear}.\\[2pt] 
%
{\bf\noindent$\bullet$ Valid Inequalities\\[2pt]}
\indent To further improve the computational efficiency of the problem formulation, here we introduce two families of valid inequalities.
First, for each flow $(k,s)$, the summation of  fractions of data rates on link $(i,j)$ over $p \in \mathcal{P}$ can be upper bounded by one.
This, together with \eqref{linearcons} and \eqref{zzrelcons}, implies
\begin{align}
& \sum_{p \in \mathcal{P}} r_{ij}(k,s,p) \leq z_{ij}(k),\nonumber \\[-3pt]
& \qquad \qquad\qquad~~~~\forall~(i,j)\in \mathcal{L},~\forall~k \in \mathcal{K}, ~ \forall~s \in \mathcal{F}(k) \cup \{0\}.\label{validineq1}
\end{align}
Second, substituting \eqref{nonlinearcons} into \eqref{consdelay2funs1}, we have $ r(k,s,p)\theta(k,s) \geq \sum_{(i,j) \in \mathcal{{L}}}  d_{ij}  r_{ij}(k, s, p) $, which, together with \eqref{relalambdaandx11}, further implies 
\begin{align}
&\theta(k,s) \geq \sum_{p\in \mathcal{P}}\sum_{(i,j) \in \mathcal{L}} d_{ij}r_{ij}(k,s,p),
~\forall~k \in \mathcal{K}, ~ \forall~s \in \mathcal{F}(k) \cup \{0\}.\label{validineq2}
\end{align}
Constraints \eqref{validineq1}-\eqref{validineq2} are redundant but {such constraints are necessary for the result in the coming Theorem \ref{thm1} and} can improve the solution efficiency of the coming problem formulation.
\\[2pt]
{\bf\noindent$\bullet$ New MILP Formulation and Analysis\\[2pt]}
\indent We are now ready to present the new MILP  formulation:
\begin{align}
& \min_{\substack{\boldsymbol{x},\boldsymbol{y},\\\boldsymbol{r},\boldsymbol{z},\boldsymbol{\theta}}} &  & \sum_{v \in \mathcal{V}}y_v + \sigma \sum_{(i,j)\in \mathcal{L}} \sum_{k\in \mathcal{K}} \sum_{s\in \mathcal{F}(k)\cup \{0\}} \sum_{p\in \mathcal{P}} r_{ij}(k,s,p) \nonumber \\
& ~~~{\text{s.\,t.~}}                                                                    &  & \eqref{onlyonenode}-\eqref{SFC1},~\eqref{linkcapcons1}-\eqref{validineq2}, \nonumber \\
& &&  x_{v,s}(k),~x_v(k),~y_v\in\{0,1\},\,\forall \,k\in\mathcal{K}, ~s\in \mathcal{F}(k),~v\in\mathcal{{V}}, \nonumber\\
& &&r_{ij}(k, s, p )\geq 0,~z_{ij}(k, s, p ),~z_{ij}(k)\in \{0,1\}, \nonumber\\
& && \qquad \qquad~~~\forall~(i,j)\in \mathcal{L},~k\in \mathcal{K},~s\in \mathcal{F}(k)\cup \{0\},~p \in \mathcal{P}, \nonumber\\
& &&  \theta(k,s)\geq 0,~\forall~k \in \mathcal{K},~s \in \mathcal{F}(k)\cup \{0\}.
\label{milp}
\tag{\text{MILP}}
\end{align}
\begin{theorem}
	\label{thm1}
	(i) Formulations \eqref{milp} and \eqref{minlp} are equivalent; (ii) moreover, the LP relaxation of formulation \eqref{milp} and the NLP relaxation of formulation \eqref{minlp} are also equivalent.
\end{theorem}
\noindent Theorem \ref{thm1} shows the advantage of formulation \eqref{milp} over formulation \eqref{minlp}.
More specifically, in sharp contrast to the continuous relaxation of \eqref{minlp} which is a (nonconvex) NLP problem, the continuous relaxation of  \eqref{milp} is a polynomial time solvable LP problem.
In addition, when compared with the LP relaxation of the linearization version of \eqref{minlp} (i.e., replacing \eqref{nonlinearcons} by \eqref{bilinearcons}), the LP relaxation of \eqref{milp} can provide a much stronger LP bound (which is as strong as that of the NLP relaxation of \eqref{minlp}).
Consequently, solving \eqref{milp} by MILP solvers should be much more computationally efficient, compared against solving \eqref{minlp}; see Section \ref{sect:numres} further ahead.

We remark that the efficient formulation \eqref{milp} is also a crucial step towards developing efficient algorithms for solving the network slicing problem due to the following two reasons.
First, globally solving the problem provides an important benchmark for evaluating the performance of heuristic algorithms.
Second, some decomposition algorithms (e.g., column generation \cite{Liu2017}) require to solve several small subproblems of the network slicing problem (e.g., problem with a single service), and employing \eqref{milp} in solving these subproblems is much more computationally efficient (as compared with \eqref{minlp}), making the overall performance much better.
\section{Numerical Simulation}
\label{sect:numres}

In this section, we present simulation results to illustrate the effectiveness and efficiency of the proposed formulations \eqref{milp} and \eqref{minlp}.
We choose $\sigma  = 0.005$ and $ P=2 $ in  \eqref{milp} and  \eqref{minlp}.
We use Gurobi 9.0.1 \cite{Gurobi} to solve \eqref{milp} and (the linearization version of) \eqref{minlp} with the time limit being 1800 seconds and the relative gap being $0.5 \% $.

The fish network topology in \cite{Zhang2017}, which contains 112 nodes and 440 links, including 6 cloud nodes, is employed in evaluating the performance in our experiments.
We randomly set the cloud nodes' and links' capacities to $ [50,100] $ and $ [7,77] $, respectively.
The NFV and communication delays on the cloud nodes and links are randomly selected as  $\{3,4,5,6\}$ and $\{1,2\}$, respectively; and the reliabilities of cloud nodes and links are randomly chosen in $[0.991,0.995]$ and $[0.995,0.999]$, respectively.
For each service $k$, node $S(k)$ is randomly chosen from the available nodes and node $D(k)$ is set to be the common destination node; SFC $ \mathcal{F}(k) $ contains 3 functions randomly generated from $ \{f^1, \ldots, f^4\} $; $ \lambda^k_s $'s are the data rates which are all set to be the same integer value, randomly chosen from $ [1,11] $; $ \Theta_k $ and $\Gamma_k$ are set to $ 20+(3*\text{dist}_k+\alpha) $ and $ 0.99^2{(\text{dist}'_k)}^4 $  where $ \text{dist}_k $ and $ \text{dist}'_k $ are the shortest paths between nodes $ S(k) $ and $ D(k) $ (in terms of delay and reliability, respectively) and $ \alpha $ is randomly chosen in $[0,5]$.
The above parameters are carefully selected to ensure that the constraints in formulations \eqref{milp} and \eqref{minlp} are neither too tight nor too loose.
For each fixed number of services, we randomly generate 100 problem instances and the results reported below are averages over the 100 instances.
\\[2pt]
\begin{figure}[t]
	\centering
	\captionsetup{belowskip=-4pt}
	\subfigure[]{
		\includegraphics[height=1.2in]{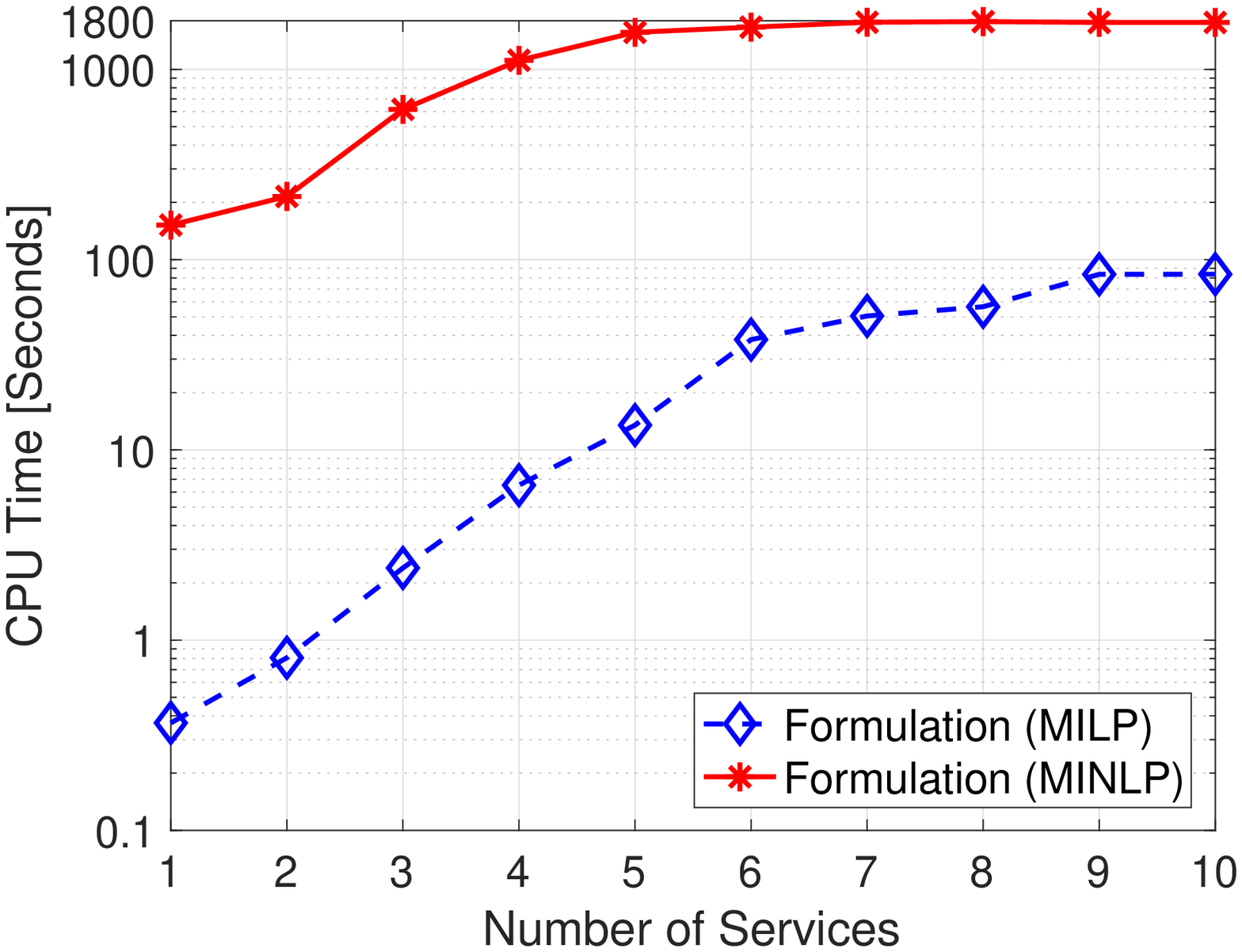}
		\label{cputime}
	}
	\subfigure[]{
		\includegraphics[height=1.2in]{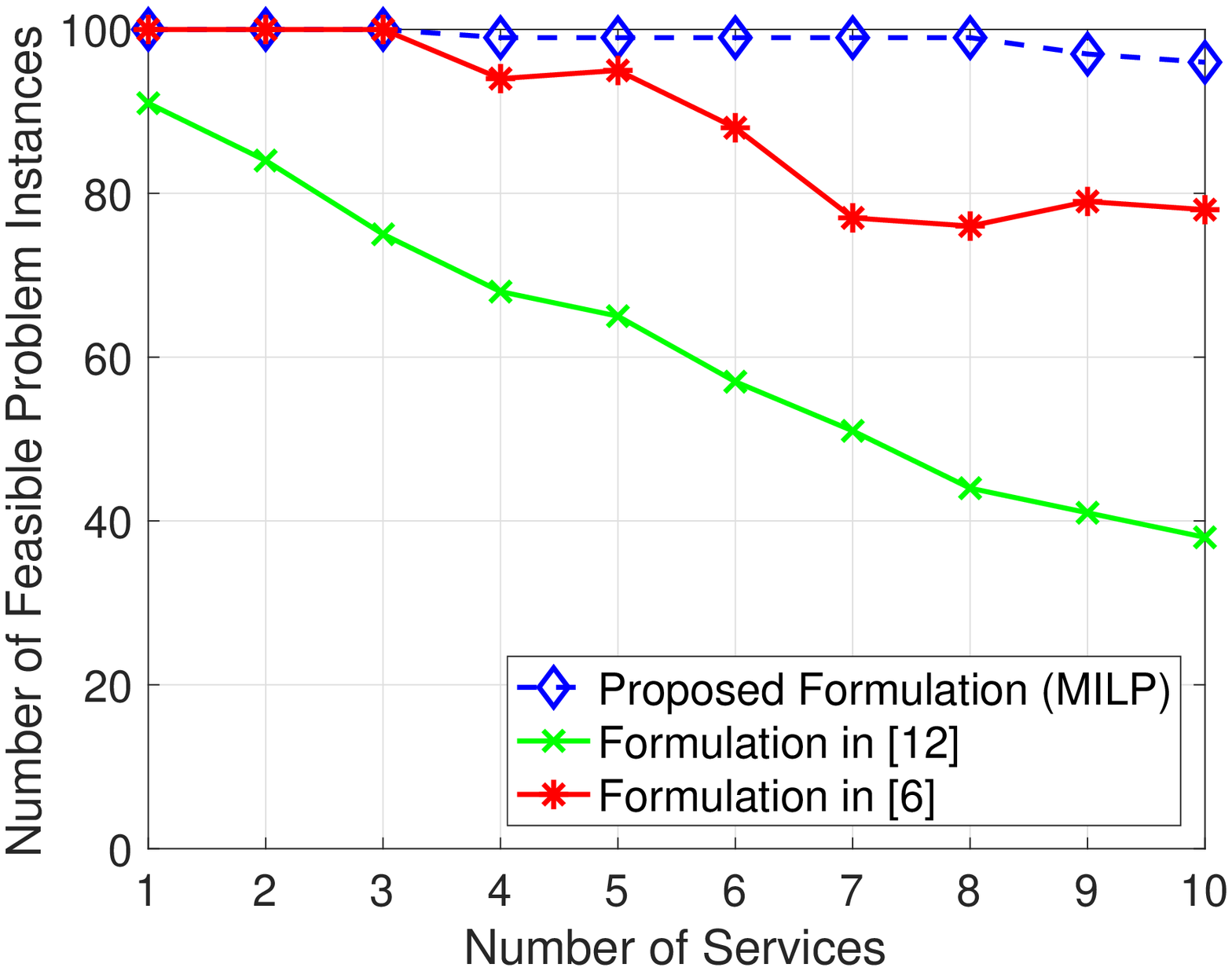}
		\label{nfeas}
	}
\vspace{-1\baselineskip}
	\caption{{Left: the average CPU time taken by solving formulations \eqref{milp} and \eqref{minlp}; Right: the number of feasible problem instances by solving formulation \eqref{milp} and those in \cite{Vizarreta2017} and \cite{Chen2021b}.}}
	\vspace*{-0.5cm}
\end{figure}
{\bf\noindent$\bullet$~Comparison of Proposed Formulations \eqref{milp} and \eqref{minlp} \\[2pt]}
\indent We first compare the solution efficiency of solving formulations \eqref{milp} and \eqref{minlp}.
Fig. \ref{cputime} plots the average CPU time taken by solving formulations \eqref{milp} and \eqref{minlp} versus the number of services.
From the figure, it can be clearly seen that it is much more efficient to solve \eqref{milp} than \eqref{minlp}.
In particular, in all cases, the CPU times of solving \eqref{milp} are all within 100 seconds, while that of solving \eqref{minlp} are larger than 1000 seconds when $|\mathcal{K}|\geq 4$.
We remark that when $|\mathcal{K}|\geq 4$, Gurobi  failed to solve \eqref{minlp}  within 1800 seconds in most cases.
This clearly demonstrates the advantage of our new way of formulating the flow conservation constraints \eqref{SFC0}-\eqref{SFC5} for the data rates over those in \cite{Chen2021b,Promwongsa2020} and the proposed valid inequalities \eqref{validineq1}-\eqref{validineq2}, i.e., it can effectively make the network slicing problem much more computationally solvable.
Next, we shall only use and discuss formulation \eqref{milp}.\\[2pt]
%
{\bf\noindent$\bullet$~Comparison of Proposed Formulation \eqref{milp} and Those in  \cite{Vizarreta2017} and \cite{Chen2021b}\\[2pt]}
\indent We show the effectiveness of our proposed formulation \eqref{milp} by comparing it with the two formulations in \cite{Vizarreta2017} and \cite{Chen2021b}.
Fig. \ref{nfeas} plots the number of feasible problem instances by solving the three formulations.
For the curves of the formulation in \cite{Vizarreta2017} and formulation \eqref{milp}, since the E2E reliability constraints \eqref{zzrelcons}-\eqref{E2Ereliability2} are explicitly enforced, we solve the corresponding formulation by Gurobi and count the corresponding problem instance feasible if Gurobi can return a feasible solution; otherwise it is claimed as infeasible.
As for the formulation in \cite{Chen2021b}, since it does not explicitly consider the E2E reliability constraints \eqref{zzrelcons}-\eqref{E2Ereliability2}, the corresponding curve in Fig. \ref{nfeas} is obtained as follows. 
We solve the formulation in \cite{Chen2021b} and then substitute the obtained solution into constraints \eqref{zzrelcons}-\eqref{E2Ereliability2}: if the solution satisfies constraints \eqref{zzrelcons}-\eqref{E2Ereliability2} of all services, we count the corresponding problem instance feasible; otherwise it is infeasible. 

As observed in Fig \ref{nfeas}, the proposed formulation \eqref{milp} can solve a much larger number of problem instances than that solved by the formulation in \cite{Vizarreta2017}, especially in the case where the number of services is large.
This shows the advantage of the flexibility of traffic routing in \eqref{milp}.
In addition, compared against that of solving the  formulation in \cite{Chen2021b}, the number of feasible problem instances of solving the proposed formulation \eqref{milp} is also much larger, which clearly shows the advantage of our proposed formulation over that in \cite{Chen2021b}, i.e., it has a guaranteed E2E reliability of the services.
To summarize, the results in Fig. \ref{nfeas} demonstrate that, compared with those in \cite{Vizarreta2017} and \cite{Chen2021b}, our proposed formulation can give a “better” solution.
More specifically, compared with that in \cite{Vizarreta2017}, our formulation can flexibly route the traffic flow; compared with that in \cite{Chen2021b}, our formulation is able to  guarantee the E2E reliability of all services.

\newpage

\bibliographystyle{IEEEtran}

\end{document}